\documentclass[conference, hidelinks]{IEEEtran}

\usepackage[
  letterspace=20,
  expansion=alltext,
  protrusion=alltext-nott,
  final
]{microtype}
\usepackage[all=normal, paragraphs=tight, floats=tight, mathspacing=tight]{savetrees}
\usepackage{amsmath, amsfonts, bm, color,amsthm, amssymb}
\usepackage{gensymb}
\usepackage{textcomp}
\usepackage{verbatim}
\usepackage{graphicx}
\usepackage{cite}
\usepackage{algorithm}
\usepackage{algpseudocode}

\usepackage{array}
\usepackage{blkarray}
\usepackage{tabularx,booktabs}
\usepackage{placeins}

\ifCLASSOPTIONcompsoc
 \usepackage[caption=false,font=normalsize,labelfont=sf,textfont=sf]{subfig}
\else
 \usepackage[caption=false,font=footnotesize]{subfig}
\fi
\usepackage{stfloats}
\usepackage{url}
\usepackage{hyperref}

\usepackage{enumitem}

\usepackage{flushend}

\newlist{inlineroman}{enumerate*}{1}
\setlist[inlineroman]{afterlabel=~,label=\roman*)}
\newcommand{\inlinerom}[1]{
\begin{inlineroman}
#1
\end{inlineroman}
}

\DeclareMathOperator{\Tr}{Tr}

\newcommand{\mch}[2]{
\left.\mathchoice
  {\left(\kern-0.48em\binom{#1}{#2}\kern-0.48em\right)}
  {\big(\kern-0.30em\binom{\smash{#1}}{\smash{#2}}\kern-0.30em\big)}
  {\left(\kern-0.30em\binom{\smash{#1}}{\smash{#2}}\kern-0.30em\right)}
  {\left(\kern-0.30em\binom{\smash{#1}}{\smash{#2}}\kern-0.30em\right)}
\right.}

\begin{document}
\title{Opportunistic Beamforming and Dynamic\\\vspace{-0.1cm} Scheduling for Multi-User MIMO-ISAC Systems 
\vspace{-4.5mm}}

\author{
\IEEEauthorblockN{Tharaka Perera\IEEEauthorrefmark{1},
Saman Atapattu\IEEEauthorrefmark{2}, Chathuranga Weeraddana\IEEEauthorrefmark{3}, and
Jamie Evans\IEEEauthorrefmark{1}}%
\IEEEauthorblockA{\IEEEauthorrefmark{1}Department of Electrical and Electronic Engineering, University of Melbourne, Victoria, Australia. \\
\IEEEauthorrefmark{2}Department of Electrical and Electronic Engineering, RMIT University, Melbourne, Victoria, Australia. \\
\IEEEauthorrefmark{3} Centre for Wireless Communication, University of Oulu, Oulu, Finland.\\
}
\vspace{-10mm}
}

\maketitle
\begin{abstract}
This research presents a novel framework integrating Flexible-Duplex (FlexD) and Integrated Sensing and Communications (ISAC) technologies to address the challenges of spectrum efficiency and resource optimization in next-generation wireless networks. We develop a unified system model for a dual-functional radar-communication base station with multiple-input multiple-output capabilities, enabling dynamic uplink and downlink channel allocation. The framework maximizes network throughput while maintaining radar sensing performance, subject to signal-to-clutter-plus-noise ratio (SCNR) requirements and power constraints. Given the non-convex and combinatorial nature of the resulting optimization problem, we propose an iterative algorithm that converges to a locally optimal solution.
Extensive simulations demonstrate the superiority of the proposed FlexD-ISAC framework compared to conventional half-duplex networks. Additionally, sensitivity analyses reveal the impact of SCNR requirements and power constraints on system performance, providing valuable insights for practical implementation. This work establishes a foundation for future research in dynamic, resource-efficient wireless systems that simultaneously support sensing and communication capabilities.
\end{abstract}

\vspace{-2.5mm}
\section{Introduction}
\vspace{-1mm}
The rapid advancement of wireless communication systems, particularly in vehicular networks~\cite{zhang2024optimal}, demands innovative solutions to address the growing requirements for enhanced data rates, reduced latency, and precise sensing capabilities. As networks evolve toward 6G, Integrated Sensing and Communications (ISAC) has emerged as a transformative paradigm, enabling critical vehicular applications, including vehicle-to-everything (V2X) and vehicle-to-infrastructure (V2I) communications~\cite{sturm2011waveform, hua2023optimal}. While conventional half-duplex (HD) systems offer implementation simplicity, they impose fundamental limitations on spectral efficiency and exhibit vulnerabilities to interference and security threats. To address these constraints, flexible-duplex (FlexD) architectures have been proposed, dynamically adapting to network conditions~\cite{perera2024flexd}.

ISAC represents a paradigm shift from traditional approaches where radar and communication functionalities operated independently. By proposing a unified framework, ISAC enables the\linebreak seamless integration of sensing and communication capabilities, sharing critical system resources including spectrum, antenna arrays, and power budget. This integration is particularly crucial\linebreak in vehicular networks, where precise sensing and reliable communications must coexist to ensure safe and efficient transportation systems~\cite{zhang2022time, jing2024isac}. Recent research has expanded ISAC\linebreak applications to indoor localization~\cite{yu2022location} and human activity monitoring~\cite{cui2021integrating}, with significant advances in waveform design~\cite{zhou2022integrated}, beamforming strategies~\cite{zhang2024joint}, and resource allocation mechanisms~\cite{chen2021joint}.

FlexD, on the other hand, has emerged as a promising approach to improve spectral efficiency by leveraging the inherent randomness of wireless channels. Unlike conventional duplexing schemes that rely on static resource partitioning, FlexD introduces adaptive allocation mechanisms that dynamically adjust uplink (UL) and downlink (DL) resources based on instantaneous network conditions. This flexibility is particularly valuable in vehicular networks, where rapidly changing channel conditions and diverse traffic patterns necessitate adaptive resource management. Recent advances in FlexD have explored resource allocation algorithms for interference broadcast channels~\cite{flex_net_2022}, security enhancement mechanisms~\cite{perera2024flexd}, and machine learning-based resource allocation~\cite{perera2024graph}.

Despite their individual advancements, ISAC and FlexD have largely been treated as separate research areas, leaving a gap in exploring their integration. \textit{This work represents the first effort to integrate FlexD and ISAC technologies, paving the way for a dynamic, resource-efficient framework that bridges the gap between these two research domains.} The main contributions of this work are as follows:
\begin{enumerate}[label=\roman*)]
    \item 
    We propose a novel system model that integrates the FlexD scheme into the ISAC paradigm. This model enables dynamic UL/DL channel allocation through user sub-grouping, allowing a dual-functional radar-communication (DFRC) multiple-input multiple-output (MIMO) base station (BS) to simultaneously support communication with multi-antenna-equipped users.
    \item 
    We propose an optimization framework for FlexD-enabled ISAC systems to dynamically optimize UL and DL directions, mitigating interference beyond fixed HD configurations. The strategy maximizes UL/DL sum rate while ensuring radar performance via a signal-to-clutter-plus-noise ratio (SCNR) constraint. Efficient algorithms are developed using a combination of convex and alternating optimization methods, ensuring practical implementation with guaranteed convergence.
    \item
    The proposed algorithm is rigorously validated through extensive numerical simulations and compared against traditional HD networks and other baseline methods. Furthermore, we investigate the impact of power and SCNR constraints on system throughput.
\end{enumerate}
\section{System Model}
\subsection{Network Model}
\begin{figure}[!t]
\centering
\includegraphics[width=\columnwidth]{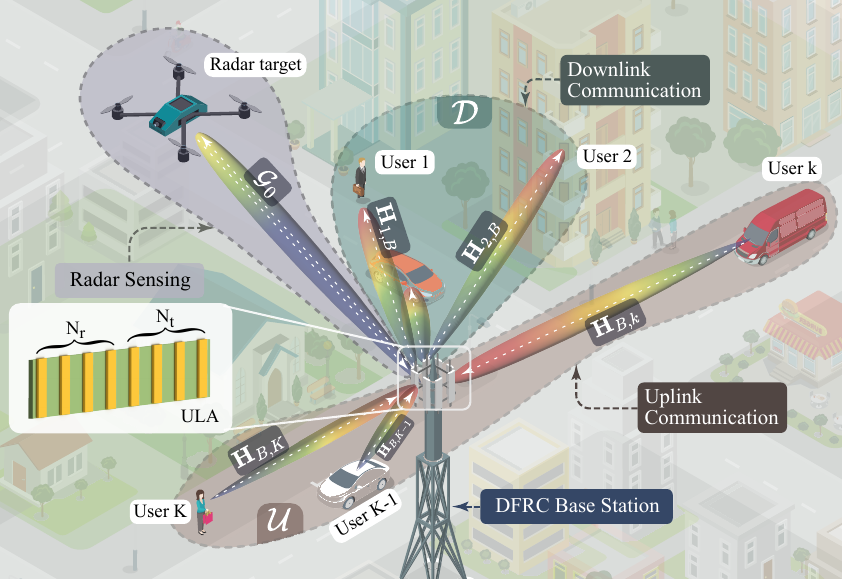}
\caption{Illustration of a multi-user MIMO FlexD-ISAC network in an urban environment, where a DFRC base station simultaneously facilitates DL/UL communications with $K$ vehicular and mobile users, while performing radar target detection.}
\label{fig:system_model}
\vspace{-3mm}
\end{figure}

As shown in Fig.~\ref{fig:system_model}, we consider a FlexD multi-user MIMO network comprising a BS with DFRC capability and $K$ users, denoted by the set $\mathcal{K} = \{1,2,\ldots,K\}$. The BS is equipped with uniform linear array (ULA) composed of $N_t$ transmit antennas for DL communication and radar operations, $N_r$ receive antennas for UL communication and radar reflection reception~ \cite{shi2022ula}. Each user $k \in \mathcal{K}$ employs $L_k$ antennas. The spatial separation of antennas at the BS enables effective mitigation of self-interference arising from DL transmission~\cite{ouyang2022performance}.

The BS concurrently performs three operations: \inlinerom{\item DL communication, \item UL communication, and \item radar sensing of a target positioned at angle $\theta_0$.} To enhance operational efficiency, the BS employs the same symbol sequence for both DL communication and radar sensing functions~\cite{bazzi2023outage}.

\subsection{FlexD Communication Model}
The FlexD framework enables dynamic allocation of communication directions by adaptively switching users between UL and DL modes. At any given time slot, the network users are partitioned into disjoint sets $\mathcal{D}$ and $\mathcal{U}=\mathcal{K}\setminus\mathcal{D}$, representing DL and UL users, respectively.

\subsubsection{Uplink Communications}
The transmit signal of user $k$ at time step $t$ is given by $\bm{x}_{B,k}[t] = \bm{V}_{B,k} \bm{s}_{B,k}[t] \in \mathbb{C}^{L_k \times 1}$, where $\bm{V}_{B,k} \in \mathbb{C}^{L_k \times T_{B,k}}$ denotes the transmit beamformer and $\bm{s}_{B,k} \in \mathbb{C}^{T_{B,k} \times 1}$ represents the unit-power signal vector, i.e., $\mathbb{E}[\bm{s}_{B,k}{\bm{s}_{B,k}}^{H}] = \bm{I}$. Here, $T_{B,k}$ denotes the length of the signal vector from user $k$ to BS.
The received signal at the BS from user $k$ is expressed as
\begin{equation}
    \bm{y}_{B,k}[t] = \bm{H}_{B,k}\bm{x}_{B,k}[t] + \sum_{i \in \mathcal{U}\setminus k} \bm{H}_{B,i}\bm{x}_{B,i}[t] + \bm{n}_{B}[t],
\end{equation}
where $\bm{H}_{B,k} \in \mathbb{C}^{N_r \times L_k}$ represents the flat Rayleigh fading channel from user $k$ to BS and $\bm{n}_B[t] \sim \mathcal{CN}(0, \sigma^2_B \bm{I})$ represents independent and identically distributed (i.i.d.) additive white Gaussian noise (AWGN) at the BS. The achievable UL communication rate at the BS from user $k$ is given by
\begin{align}
    C_{B,k} =& \log\, \det\biggl[\bm{I} + \bm{H}_{B,k}\bm{V}_{B,k}\bm{V}_{B,k}^H \bm{H}_{B,k}^H\nonumber\\
    &\biggl(\sum_{i \in \mathcal{U}\setminus k}\bm{H}_{B,i}\bm{V}_{B,i}\bm{V}_{B,i}^H\bm{H}_{B,i}^H+\sigma^2_B\bm{I}\biggr)^{-1} \Biggr].
\end{align}

\subsubsection{Downlink Communications}
The aggregate transmit signal vector from the BS is formulated as $\bm{z}[t] = \sum_{j \in \mathcal{D}} \bm{V}_{j,B} \bm{s}_{j,B}[t] \in \mathbb{C}^{N_t \times 1}$, where $\bm{V}_{j,B} \in \mathbb{C}^{N_t \times T_{j,B}}$ represents the transmit beamformer and $\bm{s}_{j,B} \in \mathbb{C}^{T_{j,B} \times 1}$ denotes the unit-power DL symbol sequence. $T_{j,B}$ denotes the length of the signal vector from BS to user $j$. Under the assumption of statistical independence of data symbols across different users, this transmit signal $\bm{z}[t]$ serves dual purposes: communication and radar sensing. At the receiver side, the signal observed by user $k$ can be expressed as
\begin{equation}
    \bm{y}_{k,B}[t] = \bm{H}_{k,B}\bm{z}[t] + \bm{n}_{k}[t],
\end{equation}
where $\bm{n}_k[t] \sim \mathcal{CN}(0, \sigma^2_k \bm{I})$ represents i.i.d. AWGN at user $k$. The achievable DL communication rate at user $k$ from BS is given by
\vspace{-4mm}
\begin{align}
    C_{k,B} =& \log\, \det\biggl[\bm{I} + \bm{H}_{k,B}\bm{V}_{k,B}\bm{V}_{k,B}^H \bm{H}_{k,B}^H\nonumber\\
    &\biggl(\sum_{j \in \mathcal{D}\setminus k}\bm{H}_{k,B}\bm{V}_{j,B}\bm{V}_{j,B}^H\bm{H}_{k,B}^H+\sigma^2_k\bm{I}\biggr)^{-1} \Biggr].
\end{align}
\vspace{-5mm}
\subsection{Sensing Model}
The proposed DFRC system utilizes the transmit signal $\bm{z}[t]$ for both communication and radar sensing purposes. The radar sensing operations are performed within the same coherent time interval as signal transmission. The composite received sensing signal at time step $t$ at the BS is characterized by
\begin{equation}
    \bm{r}[t] = \bm{G}_0 \bm{z}[t] \hspace{-0.2mm}+\hspace{-1mm} \sum_{m \in \mathcal{M}}\hspace{-1mm}\bm{G}_m \bm{z}[t] + \sum_{i \in \mathcal{U}} \bm{H}_{B,i} \bm{x}_{B,i}[t] + \bm{n}_B[t],
\end{equation}
where $\mathcal{M}$ denotes the set of clutter elements in the environment. The channel matrices $\bm{G}_0$ and $\bm{G}_m$ represent the target and clutter reflection channels, respectively, defined as
$\bm{G}_0 = \beta_0 \bm{a}_r(\theta_{0}) \bm{a}_t(\theta_{0})^H$ and 
$\bm{G}_m = \beta_{m} \bm{a}_r(\theta_{m}) \bm{a}_t(\theta_{m})^H$,
where $\beta_0$ and $\beta_m$ represent the complex reflection coefficients for the target and clutter elements, respectively. The receive and transmit steering vectors are given by $\bm{a}_r(\theta) = \frac{1}{\sqrt{N_r}}[1, \ldots, e^{j2\pi \frac{d_r}{\lambda}(N_r - 1)\sin(\theta)}]^T $ and $\bm{a}_t(\theta) = \frac{1}{\sqrt{N_t}}[1, \ldots, e^{j2\pi \frac{d_t}{\lambda}(N_t - 1)\sin(\theta)}]^T$. Here, $\lambda$ denotes the wavelength, while $d_r$ and $d_t$ represent the inter-element spacing of receive and transmit antenna arrays, respectively. The intended target is positioned at angle $\theta_0$, and the $m$th clutter element is located at angle $\theta_m$.
The third term incorporates the interference from UL users on the radar sensing signal. To evaluate radar performance, we focus on the signal-to-clutter-plus-noise ratio (SCNR), as detection probability exhibits monotonic increase with SCNR \cite{khawar2015target}.
For a receive beamformer $\bm{q} \in \mathbb{C}^{N_r \times 1}$, the SCNR is expressed as
\begin{equation}\label{eq:gamma_expression}
    \gamma_r(\bm{q}) = \mathbb{E}\left[\frac{|\beta_0 \bm{q}^H \bm{A}(\theta_0) \bm{z}|^2}{\bm{q}^H \bm{R} \bm{q}}\right],
\end{equation}
where $\bm{A}(\theta) = \bm{a}_r(\theta) \bm{a}_t(\theta)^H \in \mathbb{C}^{N_r \times N_t}$ and the interference-plus-clutter covariance matrix $\bm{R} \in \mathbb{C}^{N_r \times N_r}$ is given by
\begin{align}\label{eq:clutter_plus_noise}
    \bm{R} = \sum_{m\in \mathcal{M}}|\beta_{m}|^2 \bm{A}&(\theta_{m})\left[\sum_{j \in \mathcal{D}}\bm{V}_{j,B}{\bm{V}_{j,B}}^H\right] \bm{A}^H(\theta_{m})\nonumber\\
    &+\sum_{i \in \mathcal{U}} \bm{H}_{B,i} \bm{V}_{B,i} {\bm{V}_{B,i}}^H {\bm{H}_{B,i}}^H+\sigma_B^2\bm{I}.
\end{align}
By determining the optimal minimum variance distortionless response (MVDR) beamformer $\bm{q}^*$~\cite{capon1969high} and substituting it into \eqref{eq:gamma_expression}, we obtain
\begin{align}\label{eq:scnr}
    \gamma_r(\bm{q}^*) &= |\beta_0|^2 \sum_{j \in \mathcal{D}}\text{Tr}\left(\bm{V}_{j,B} {\bm{V}_{j,B}}^H \bm{\Theta}\right),
\end{align}
where $\bm{\Theta} = \bm{A}^H(\theta_0) \bm{R}^{-1} \bm{A}(\theta_0) \in \mathbb{C}^{N_t \times N_t}$.

\subsection{Optimization Problem}
The network aims to maximize the sum data rate across both UL and DL communications while satisfying power constraints and maintaining a specified minimum radar SCNR.
The optimization problem can be formally expressed as
\begin{subequations}
\begin{alignat}{2}
    \hspace{-1mm}\underset{\{\bm{V}_{B,k}, \bm{V}_{k,B}, \mathcal{D}\}}{\max} \quad & \sum_{i \in \mathcal{K}\setminus\mathcal{D}} C_{B,i} + \sum_{j \in \mathcal{D}} C_{j,B},\label{eq:objective_fn} \\
    \text{ s.t }\quad & |\beta_0|^2 \sum_{j \in \mathcal{D}} \Tr\left( \bm{V}_{j,B} {\bm{V}_{j,B}}^H \bm{\Theta} \right) &&\geq \gamma_{\text{min}},\label{eq:radar_constraint} \\
    & \sum_{j \in \mathcal{D}} \Tr\left( \bm{V}_{j,B} {\bm{V}_{j,B}}^H\right) &&\leq P^{\text{(BS)}}_{\max},\label{eq:bs_power_constraint} \\
    &  \Tr\left( \bm{V}_{B,k} {\bm{V}_{B,k}}^H\right) &&\leq P^{(k)}_{\max}\;\;; \forall k,\label{eq:user_power_constraint}
\end{alignat}
\end{subequations}
where \eqref{eq:radar_constraint}, \eqref{eq:bs_power_constraint}, and \eqref{eq:user_power_constraint} represent radar SCNR constraint, power constraint of the BS, and power constraints of user devices, respectively. When the radar SCNR constraint cannot be satisfied under the given power constraints and channel conditions, the problem becomes infeasible, resulting in system outage.
Given the non-convex and combinatorial nature of this optimization problem, we employ a transformation to an equivalent sum mean square error (MSE) minimization problem, following the approach in \cite{shi2011iteratively}. The reformulated problem is expressed as
\begin{align}\label{eq:opt_wmmse}
    \underset{\{\bm{V},\bm{U},\bm{W}, \mathcal{D}\}}{\min} \quad & \sum_{i \in \mathcal{K}\setminus\mathcal{D}} \Tr \left( \bm{W}_{B,i} \bm{E}_{B,i}\right) - \log\:\det\left(\bm{W}_{B,i} \right) \\
    &+ \sum_{j \in \mathcal{D}} \Tr \left( \bm{W}_{j,B} \bm{E}_{j,B} \right) - \log\:\det\left(\bm{W}_{j,B} \right),\nonumber \\
    \text{ s.t }\quad &\eqref{eq:radar_constraint},~\eqref{eq:bs_power_constraint},~\eqref{eq:user_power_constraint}.\nonumber
\end{align}
In this formulation, $\bm{V}\in\{\bm{V}_{B,k}, \bm{V}_{k,B}\}$, $\bm{U}\in\{\bm{U}_{B,k}, \bm{U}_{k,B}\}$, and $\bm{W}\in\{\bm{W}_{B,k}, \bm{W}_{k,B}\}$ collectively represent both UL and DL variables. The MSE matrices $\bm{E}_{ji}$ are defined as
\begin{align}
    &\bm{E}_{j,i} = (\bm{I} - {\bm{U}_{j,i}}^H \bm{H}_{j,i} \bm{V}_{j,i})(\bm{I} - {\bm{U}_{j,i}}^H \bm{H}_{j,i} \bm{V}_{j,i})^H \nonumber\\
    &+ \hspace{-4mm}\sum_{(m,\ell) \neq (j,i)} \hspace{-4mm}{\bm{U}_{j,i}}^H \bm{H}_{j,\ell} \bm{V}_{m,\ell} {\bm{V}_{m,\ell}}^H {\bm{H}_{j,\ell}}^H \bm{U}_{j,i} + \sigma_j^2 \bm{U}_{j,i}^H \bm{U}_{j,i},
\end{align}
where the weight matrices $\bm{W}_{B,i} \succeq 0$ and $\bm{W}_{j,B} \succeq 0$ correspond to UL and DL communications, respectively, and $\bm{U}_{B,i}$ and $\bm{U}_{j,B}$ denote their respective receive beamformers.

\section{Beamformer and User Partition Design\texorpdfstring{\\for FlexD Communications}{}}
\subsection{Uplink Beamformer Design}
With given DL beamformers, the optimal UL transmit beamformers can be written as
\vspace{-1mm}
\begin{align}\label{eq:ul_tr_beamformer}
    &\bm{V}_{B,k} =\biggl({\bm{H}_{B,k}^H} \bm{U}_{B,k}\bm{W}_{B,k} {\bm{U}_{B,k}^H}\bm{H}_{B,k}+
    \sum_{j \in \mathcal{D}}{\bm{H}_{jk}^H}\bm{U}_{j,B} \bm{W}_{j,B}\nonumber\\ &\hspace{2mm}\times {\bm{U}_{j,B}^H} \bm{H}_{jk} + \lambda \bm{I} +  \mu \biggl[ |\beta_0|^2{\bm{H}_{B,k}^H} \bm{R}^{-1}\bm{A}(\theta_0)\sum_{j \in \mathcal{D}} \bm{V}_{j,B} \bm{V}_{j,B}^H\nonumber\\
    &\qquad\times {\bm{A}(\theta_0)}^H \bm{R}^{-1} \bm{H}_{B,k} \biggl] \biggl)^{-1}\bm{H}_{B,k}^H \bm{U}_{B,k}\bm{W}_{B,k},
    \vspace{-1mm}
\end{align}
where $\lambda$ and $\mu$ represent the Lagrangian multipliers associated with the power and radar constraints, respectively.
\begin{IEEEproof}
The optimization problem's Lagrangian incorporating user power and radar constraints follows
\vspace{-1mm}
\begin{align}
    \mathcal{L}&\left( \bm{V}_{B,k}, \lambda, \mu \right) = \sum_{i \in \mathcal{K}\setminus\mathcal{D}} \Tr \left( \bm{W}_{B,i} \bm{E}_{B,i}\right) - \log\:\det\left(\bm{W}_{B,i} \right)\nonumber \\
    &+ \sum_{j \in \mathcal{D}} \Tr \left( \bm{W}_{j,B} \bm{E}_{j,B} \right) - \log\:\det\left(\bm{W}_{j,B} \right) \nonumber \\ &\quad+ \lambda\left[ \Tr\left( \bm{V}_{B,k} {\bm{V}_{B,k}}^H\right) - P^{(k)}_{\max}\right]\nonumber\\ &\quad\quad- \mu \left[ |\beta_0|^2 \sum_{j \in \mathcal{D}} \Tr\left( \bm{V}_{j,B} {\bm{V}_{j,B}}^H \bm{\Theta} \right) - \gamma_{\text{min}}\right].
    \vspace{-1mm}
\end{align}
Taking the first-order derivative with respect to $\bm{V}_{B,k}$ and setting to zero yields \eqref{eq:ul_tr_beamformer}. The non-linearity in $\bm{R}^{-1}$ which contains $\bm{V}_{B,k}$ terms is addressed through iterative approximation, where $\bm{R}^{(t)}$ from the previous iteration is used to compute $\bm{V}_{B,k}^{(t+1)}$. This linearization technique generates a sequence of feasible beamformers that converge to a stationary point of \eqref{eq:opt_wmmse} under mild initialization conditions.
\end{IEEEproof}
Through first-order optimality conditions of \eqref{eq:opt_wmmse} with respect to $\bm{U}_{B,k}$ and $\bm{W}_{B,k}$, we can derive the receive beamformer and the weight matrices, respectively, as
\vspace{-2mm}
\begin{equation}\label{eq:beamformer_receive_ul}
    \hspace{-1.5mm}\bm{U}_{B,k} = \left( \sum_{i \in \mathcal{U}} \bm{H}_{B,i}\bm{V}_{B,i}\bm{V}_{B,i}^H\bm{H}_{B,i}^H+\sigma_B^2\bm{I}\right)^{-1}\hspace{-2mm}\bm{H}_{B,k}\bm{V}_{B,k},
    \vspace{-2mm}
\end{equation}
\begin{equation}\label{eq:weight_ul}
        \text{and}\qquad\bm{W}_{B,k} = \left( \bm{I} - \bm{U}_{B,k}^H\bm{H}_{B,k}\bm{V}_{B,k} \right)^{-1}.
\end{equation}

\begin{figure*}[!b]%
\vspace*{-3mm}
\centering
\begin{minipage}{0.32\textwidth}
\centering
\includegraphics[width=\textwidth]{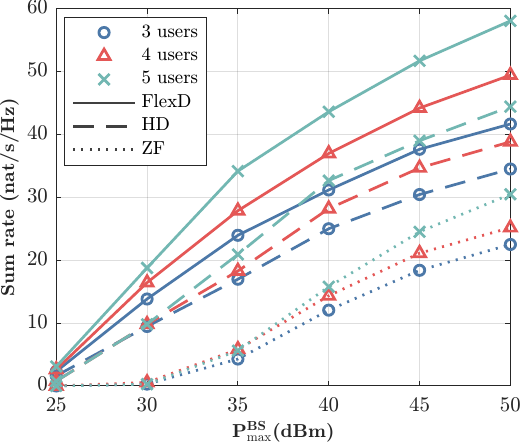}
\vspace*{-5mm}
\caption{Sum rate against maximum BS power for different network sizes when $P_{\max}^{(k)}=30$\;dBm $;\forall k$}
\label{fig:bs_power}
\end{minipage}\hfill
\begin{minipage}{0.32\textwidth}
\centering		\includegraphics[width=\textwidth]{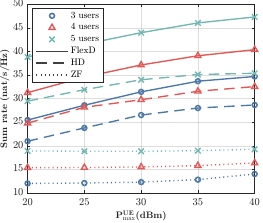}
\vspace*{-5mm}
\caption{Sum rate against maximum BS power for different network sizes when $P_{\max}^{(BS)}=40$\;dBm.}
\label{fig:user_power}
\end{minipage}\hfill
\begin{minipage}{0.32\textwidth}
\centering
\includegraphics[width=\textwidth]{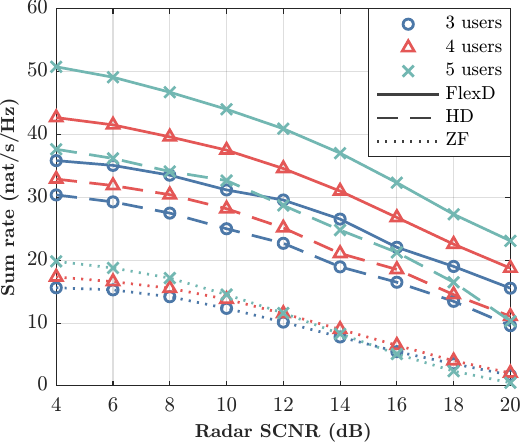}
\vspace*{-5mm}
\caption{Trade-off of communication throughput and radar SCNR constraint for different network sizes.}
\label{fig:scnr}
\end{minipage}
\end{figure*}
\vspace{-2mm}
\subsection{Downlink Beamformer Design}
For given UL beamformers, the optimal DL transmit beamformers can be expressed as
\begin{align}\label{eq:dl_tr_beamformer}
    \bm{V}_{k,B} \hspace{-.5mm}=\hspace{-.5mm}\biggl(\sum_{j \in \mathcal{D}}{\bm{H}_{j,B}^H} \bm{U}_{j,B} \bm{W}_{j,B} &{\bm{U}_{j,B}^H} \bm{H}_{j,B} + \lambda \bm{I} -  \mu |\beta_0|^2 \bm{\Theta} \biggl)^{-1}\nonumber\\
    &\times{\bm{H}_{B,k}}^H \bm{U}_{B,k}\bm{W}_{B,k}.
\end{align}
\begin{IEEEproof}
Proof is similar to that of \eqref{eq:ul_tr_beamformer} except BS power constraint and radar constraint are incorporated to the Lagrangian. Non-linearity in $\bm{V}_{k,B}^{(t+1)}$ is eliminated using $\bm{R}^{(t)}$ from the previous iteration.
\end{IEEEproof}
Similarly, the corresponding receive beamformer and weight matrices are obtained respectively as
\vspace{-2mm}
\begin{equation}\label{eq:beamformer_receive_dl}
    \hspace{-1.5mm}\bm{U}_{k,B} = \left( \sum_{j \in \mathcal{D}} \bm{H}_{k,B}\bm{V}_{j,B}\bm{V}_{j,B}^H\bm{H}_{k,B}^H+\sigma_k^2\bm{I}\right)^{-1}\hspace{-2mm}\bm{H}_{k,B}\bm{V}_{k,B},
    \vspace{-2mm}
\end{equation}
\begin{equation}\label{eq:weight_dl}
    \text{and}\qquad\bm{W}_{k,B} = \left( \bm{I} - \bm{U}_{k,B}^H\bm{H}_{k,B}\bm{V}_{k,B} \right)^{-1}.
\end{equation}
\setlength{\textfloatsep}{6pt}
\setlength{\floatsep}{5pt}
\begin{algorithm}[!t]\label{alg:wmmse}
\small
\caption{FlexD-ISAC Sum rate maximization}
\begin{algorithmic}[1]
\State Initialize $\bm{V}_{B,k}$ and $\bm{V}_{k,B}$ such that~\eqref{eq:radar_constraint},~\eqref{eq:bs_power_constraint}, and~\eqref{eq:user_power_constraint} are satisfied using \textbf{Algorithm 2}.
\State $\bm{V}_{B,k}^* \gets \bm{0}; \bm{V}_{k,B}^* \gets \bm{0}$; $\mathcal{D}^* \gets \varnothing$; $rate^* \gets 0$
\For{$\mathcal{D} \gets \operatorname{PatternSearch}(\mathcal{K})$}
\State $r \gets 0$
\Repeat
\State $\bm{W}_{B,k}^\prime \gets \bm{W}_{B,k}$; $\bm{W}_{k,B}^\prime \gets \bm{W}_{k,B}$
\State Update $\bm{U_{B,k}}$ and $\bm{U_{k,B}}$ using \eqref{eq:beamformer_receive_ul}, \eqref{eq:beamformer_receive_dl}
\State Update $\bm{W}_{B,k}$ and $\bm{W}_{k,B}$ using \eqref{eq:weight_ul}, \eqref{eq:weight_dl}
\State Update $\bm{V}_{B,k}$ and $\bm{V}_{k,B}$ using \eqref{eq:ul_tr_beamformer}, \eqref{eq:dl_tr_beamformer}
\State $rate \gets {\displaystyle \sum_{i \in \mathcal{K}\setminus\mathcal{D}}} \log\:\det\bm{W}_{B,i} + {\displaystyle\sum_{j \in \mathcal{D}}} \log\:\det\bm{W}_{j,B}$
\Until $\big\rvert r -\bigl(\sum_{i \in \mathcal{U}} \log\:\det\bm{W}_{B,i}^\prime + \sum_{j \in \mathcal{D}} \log\:\det\bm{W}_{j,B}^\prime\bigr)\big\rvert \leq \epsilon$
\If{$rate^* \leq r$}
\State $\bm{V}^*_{k,B} \gets \bm{V}_{k,B}$; $\bm{V}^*_{B,k} \gets \bm{V}_{B,k}$; $\mathcal{D}^* \gets \mathcal{D}$ 
\EndIf
\EndFor
\end{algorithmic}
\end{algorithm}

\begin{algorithm}[!t]
\small
\caption{Beamformer initialization}
\begin{algorithmic}[1]
\State Initialize $\bm{V}_{k,B}$ and $\bm{V}_{B,k}$ using the pseudo-inverse of the  channel matrices (ZF solution)
\State $\bm{V}_{B,k} \gets \frac{P_{\max}^{(k)}}{\Delta 
\| \bm{V}_{B,k} \|}\bm{V}_{B,k}$
\State $\gamma_r \gets 0$
\Repeat
\State $\gamma_r^\prime \gets \gamma_r$
\State Update $\bm{R}$ using \eqref{eq:clutter_plus_noise}
\State $\bm{\Theta} \gets \bm{A}^H(\theta_0) \bm{R}^{-1} \bm{A}(\theta_0)$
\State Compute the eigendecomposition of $\bm{\Theta} = \bm{Q}\bm{\Lambda}\bm{Q}^H$
\State Take top $N_r$ eigenvectors and eigenvalues as $\bm{Q}^\prime$ and $\bm{\Lambda}^\prime$
\State $\bm{\Lambda} \gets \sqrt{\frac{P_{\max}^{\text{(BS)}}}{|\mathcal{K}|}} \frac{\bm{\Lambda}}{ \| \bm{\Lambda} \|}$
\State $\bm{V}_{k,B} \gets \bm{Q} \operatorname{diag}(\bm{\Lambda})$
\State Calculate $\gamma_r$ using \eqref{eq:scnr}
\Until $|\gamma_r - \gamma_r^\prime| \leq \epsilon$
\end{algorithmic}
\end{algorithm}

\vspace{-6mm}
\subsection{User Partition Design and Optimization Algorithm}
The proposed solution employs a two-stage optimization strategy where \inlinerom{
\item for a given set of users assigned to DL transmission (combinatorial variable $\mathcal{D}$), the algorithm iteratively optimizes the continuous variables representing the beamformers (${\bm{V}, \bm{U}, \bm{W}}$) until convergence, and
\item upon achieving a stationary point in the inner iteration, the algorithm explores alternative subsets $\mathcal{D} \subseteq\mathcal{K}$ to determine the optimal allocation of users between UL and DL transmission modes.}
The solutions to the beamformer optimization problems in equations \eqref{eq:ul_tr_beamformer} and \eqref{eq:dl_tr_beamformer} that satisfy the Karush-Kuhn-Tucker (KKT) conditions are obtained through the modified Powell method~\cite{yuan2000review}.
To address the computational complexity of exhaustive search over $2^K$ possible UL/DL combinations and the challenges associated with discrete optimization variables, we employ a direct search algorithm adapted from~\cite{Dayarathna2021}. The complete optimization procedure is detailed in Algorithm 1.

To ensure feasibility, the transmit beamformers must be properly initialized to satisfy all constraints. Algorithm 2 presents a systematic initialization procedure that consists of the steps: \inlinerom{
\item initialize transmit beamformers using the zero-forcing (ZF) solution by computing the pseudo-inverse of the channel matrix,
\item scale the UL beamformers to comply with maximum transmit power constraints, further reducing their power by a user-defined factor $\Delta$ to minimize interference with the radar signal, and
\item iteratively enhance the SCNR using techniques analogous to spatial multiplexing in MIMO systems.
}

The computational complexity of Algorithm 1 varies with the cardinality of the UL set $\mathcal{U}$ and DL set $\mathcal{D}$. Assuming equal cardinality and denoting the number of Lagrange multiplier estimation iterations as $\mathcal{I}$, the time complexity is
$\mathcal{O}\bigl(\mathcal{I}K^5\max\{N_t, N_r, L_k\}^3\bigl)$, where a $K^3$ complexity originates from the pattern search algorithm. Under similar assumptions, the time complexity of Algorithm 2 is $\mathcal{O}\bigl(KL_k+N_r^3+ N_t^3\bigl)$.
\vspace{-1mm}
\section{Numerical Results}

This section presents a comparative analysis of our proposed framework against established baselines and evaluates the optimization approach through comprehensive simulations. 
The simulation environment consists of a FlexD-ISAC network deployed over a $1 \times 1\;\text{km}^2$ area with randomly distributed users. The BS employs a ULA with $N_t=6$ transmit and $N_r=4$ receive antennas, spaced half-wavelength apart. Each user has $L_k=4$ antennas. Both BS-user and user-user channels are flat Rayleigh fading. 
The radar target is at $45\degree$, with clutter sources at $0\degree$ and $90\degree$. Reflection coefficients are based on free-space path loss with a $10$;dB antenna gain. For comparative analysis, we implement HD and ZF baseline strategies, maintaining equivalent user distribution between UL and DL sets, with uniform temporal resource allocation.

Fig.~\ref{fig:bs_power} illustrates the relationship between average communication throughput and maximum BS transmit power.
The proposed FlexD algorithm exhibits monotonically increasing throughput with elevated power levels, with higher user densities yielding better performance due to enhanced diversity gain. The FlexD approach consistently outperforms baseline methods across all power levels. At $P_{\max}^{(BS)}=40\;\text{dBm}$, in 5-user network, FlexD achieves performance gains of $33\%$ and $176\%$ compared to HD and ZF strategies, respectively, with this performance differential widening at higher power levels. Similarly, Fig.~\ref{fig:user_power} illustrates the relationship between communication throughput and maximum user power. When user power increases from $20\;\text{dBm}$ to $40\;\text{dBm}$ (a factor of 100), throughput increases by approximately $10\;\text{nat/s/Hz}$ across all user counts. This limited improvement stems from radar sensing constraints, where excessive user power generates interference that compromises sensing performance.

\begin{figure}[!t]
\centering
\includegraphics[width=0.65\columnwidth]{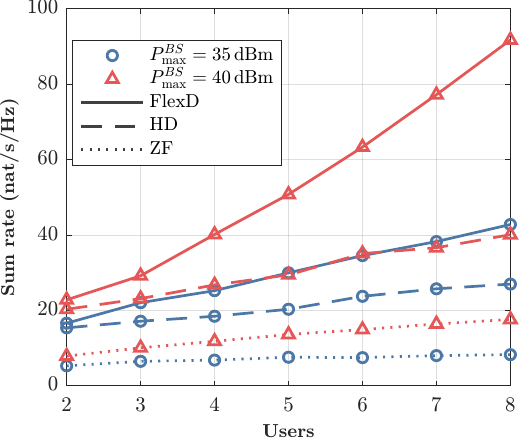}
\vspace*{-3mm}
\caption{Communication throughput against the network size comparison for different $P^{\text{(BS)}}_{\max}$ values at $P^{\text{(k)}}_{\max}=30\;\text{dBm}\;;\forall k$.}
\label{fig:users}
\vspace{-1mm}
\end{figure}
Fig.~\ref{fig:scnr} reveals the fundamental trade-off between communication throughput and SCNR constraints. At low SCNR requirements, throughput stabilizes, indicating a non-binding constraint regime. Conversely, higher SCNR requirements necessitate increased energy allocation toward radar targeting, resulting in throughput degradation. For instance, in 4-user case, doubling the sensing performance requirement from $10\;\text{dB}$ to $13\;\text{dB}$ results in a $12.6\%$ decrease in throughput.

Fig.~\ref{fig:users} demonstrates throughput performance as a function of network user count. The results highlight FlexD networks' ability to effectively leverage diversity gain for performance enhancement. The performance advantage of FlexD over baseline approaches increases with user density. At $P^{\text{(BS)}}_{\max}=40\;\text{dBm}$, FlexD achieves more than double the throughput in a 7-user network while maintaining equivalent hardware and signal processing complexity.

\vspace{-1mm}
\section{Conclusion}
\vspace{-1mm}
This study presents a novel integration of ISAC and FlexD technologies to enhance spectral efficiency under dynamic wireless channel conditions. We developed an iterative optimization algorithm that maximizes system throughput while maintaining radar SCNR constraints. Numerical results demonstrate the superior performance of our proposed framework compared to HD and ZF baselines. Additionally, we conducted a detailed analysis of power and SCNR constraint effects on the optimization approach. Future research directions include developing neural network-based approaches to reduce direction selection complexity and formulating efficient algorithms for Lagrangian multiplier determination in transmit beamformer design.

\vspace{-1mm}
\renewcommand{\baselinestretch}{0.95}
\bibliographystyle{IEEEtran}
\bibliography{IEEEabrv,refs}
\end{document}